\documentclass{article}
\newcommand{\bfr}{\begin{flushright}}
\newcommand{\efr}{\end{flushright}}
 
\begin{document}
% \eqsec  % uncomment this line to get equations numbered by (sec.num)
\title{Finite Temperature and Density Effect on Symmetry Breaking by
Wilson Loops
%\thanks{Presented at ...}%
% you can use '\\' to break lines
}
\author{
Kiyoshi Shiraishi\\
Department of Physics, Tokyo Metropolitan University,\\ Setagaya, Tokyo,
158 Japan
}
\date{Zeitschrift f\"ur Physik {\bf C35} (1987) pp. 37--42
}
\maketitle
\begin{abstract}
A finite temperature and density effect of Wilson loop elements on
non-simply connected space is investigated in the model suggested by
Hosotani. Using one-loop calculations it is shown that the value of an
``order parameter'' does not shift as the temperature grows. We find
that finite density effect is of much importance for restoration of
symmetry.
\end{abstract}
%\PACS{}

%%%%%%%%%%%%%%%%%%%%%%%%%%%%%%%%%%%%%%%%%%%%%%%%%%%%%%%%%%%%
\section{Introduction}
%%%%%%%%%%%%%%%%%%%%%%%%%%%%%%%%%%%%%%%%%%%%%%%%%%%%%%%%%%%%
For the unification of interactions between elementary particles there
has been a renewed interest in higher-dimensional theories. In
particular recent development of superstring theories \cite{1} attracts
much attention of many particle physicists. The development has been
prompted by the discovery of anomaly-free nature of the Green-Schwartz
string theory \cite{2}. 
Soon heterotic string was constructed by Gross et al. \cite{3};
that is the string theory with gauge symmetry group
$E_8 \times E_8$ or $O(32)/Z_2$. The gauge group $E_8 \times E_8$ is
sufficient for incorporating the observed fundamental gauge
interactions. Thus many string phenomenologists investigate the
so-called ``superstring-inspired models'' \cite{4} and there has been
detailed discussion on them.

Contrary to the ``pure'' Kaluza-Klein concept \cite{5}, the gauge
symmetry must be broken, rather than be generated, to the one favored
phenomenologically when extra dimensions are compactified. Symmetry
breaking is expected to be caused by Wilson loop elements on non-simply
connected internal space in consistent with supersymmetry \cite{6, 7};
incidentally non-simply connected space is required to reduce the
number of fermion families \cite{6}.

Somewhat similar mechanisms have been used previously by
Hosotani \cite{8} and Toms \cite{9}. Recently Midorikawa and
Tomiya \cite{10} have shown the detailed implication of the symmetry
breaking mechanism at ``classical'' level. In the models considered by
the above authors, a constant vacuum gauge field is allowed on the
multiply-connected space; the vacuum expectation value of the gauge
field configuration plays the role of an ``order parameter''. Moreover
Hosotani has considered a simple model to show that one-loop vacuum
effect determines the order parameter \cite{8}. Evans and Ovrut have
calculated the vacuum energy splitting of gauge field vacua on various
multiply connected manifolds \cite{11}.

Apart from string theories, an interesting unified model with $SU(5)$
symmetry using Hosotani's mechanism was suggested by Svetovo\v{i} and
Khariton \cite{12}, which naturally explained the existence of light
Higgs doublet. Although there has been no realistic model, Hosotani's
mechanism may give a new aspect of symmetry breaking in the unified
model with and without supersymmetry.

On the other hand, for modern cosmologists, the phase transition in the
history of our universe is a very important issue \cite{13}. In ordinary
field theories including spontaneous symmetry breaking, the symmetry
is expected to be restored at high temperature \cite{14} or high
density \cite{15}. The finite temperature effect is also important for
an investigation of stable compactification in Kaluza-Klein
theories \cite{16}. 

In the present paper, we consider the finite temperature
and density effect on symmetry breaking mechanism proposed by Hosotani.
In Sect. II, we discuss Hosotani's model and the mass levels of the
particles obtained by compactification. In Sect. III, we consider the
effective potential at finite temperature, namely, the free energy.
Finite density effect of fermions is considered in Sect. IV. Finally,
Sect. V is devoted to discussion.

%%%%%%%%%%%%%%%%%%%%%%%%%%%%%%%%%%%%%%%%%%%%%%%%%%%%%%%%%%%%
\section{The Wilson Loop on $S^1$ and the Mass Spectrum
}
%%%%%%%%%%%%%%%%%%%%%%%%%%%%%%%%%%%%%%%%%%%%%%%%%%%%%%%%%%%%
In this paper, we concentrate our attention to Hosotani's model in
four dimensions \cite{8}. Further, for simplicity, we treat the $SU(2)$
gauge symmetry\cite{10}. 

First suppose that the spacetime topology is
$M_3 \times S^1$ where $M_3$ is a three dimensional Minkowski
space-time and $S^1$ is a circle whose circumference is $L$. The four
dimensional coordinates are represented by $(x, y)$, where $0\le y<L$.

Let us consider $SU(2)$ gauge theory in this space-time; for the
present, we take $SU(2)$ gauge fields $A_\mu$ and fundamental massless
fermions $\psi$. $S^1$ is a compact and non-simply connected space,
i.e.,
$\Pi_1(S^1)=Z$. Boundary conditions on $A_\mu$ and $\psi$ are chosen
definitely as follows: 
\begin{equation}
\left\{\begin{array}{l}
A_\mu(x, y + L)= A_\mu(x, y) \\
\psi(x, y+ L) =e^{i\delta}
\psi(x, y)\end{array}\right.\,.
\end{equation}
At this level, symmetry breaking is not observed at all.

On the multiply-connected manifold, non zero vacuum gauge configuration
is permitted. In our case, we set
\begin{equation}
\langle gLA_y\rangle=\frac{1}{2}\left[
\begin{array}{cc}\phi & 0\\ 0 & -\phi\end{array}\right]
=\frac{\phi}{2}\tau_3\,,
\end{equation}
where the vacuum gauge field has been diagonalized by using the freedom
of gauge transformations and $g$ is the gauge coupling constant. If we
just consider non-singular gauge transformation, $\langle gLA_y\rangle$
is inequivalent to the trivial vacuum gauge field $\langle
gLA_y\rangle=0$ unless
$\langle gLA_y\rangle=2\pi n \tau_3$ ($n$: integer). We can only
restrict
$0\le \phi < 4\pi$ by gauge transformations. The value of a Wilson loop
element is:
\begin{equation}
U=P\exp(-ig\int_\gamma A_y dy)=\left[
\begin{array}{cc}e^{-i\frac{\phi}{2}n} & 0\\ 0 &
e^{i\frac{\phi}{2}n}\end{array}\right]\,,
\end{equation}
where the closed path $\gamma$ winds around $S^1$ $n$ times. Note
that $U$ is regarded as an element of the additive group $Z$
($=\Pi_1(S^1)$).

On a general multiply-connected manifold $U\ne 1$ necessarily; moreover,
on certain manifolds, $U$ is taken to be an element of a finite group,
then we can tell the symmetry breaking pattern by classification of $U$.

Now in our model, in view of the three-dimensional flat space, one
finds ladders of the discrete mass levels of infinite number of fields
by Fourier expansions in terms of $y$; those masses are:
\begin{equation}
\frac{1}{L}(2\pi n+ \phi)\quad (n{\rm :~integer})
\end{equation}
which come from the expansion of gauge fields $A^1_\mu$ and $A^2_\mu$
\begin{equation}
\frac{1}{L}(2\pi n)\quad (n{\rm :~integer})
\end{equation}
which come from the expansion of $A^3_\mu$ and
\begin{equation}
\frac{1}{L}\left(2\pi n\pm\frac{\phi}{2}-\delta\right)\quad (n{\rm
:~integer})
\end{equation}
which come from the expansion of fermions.

In general, $\phi$ lifts the mass level; this effect reduces the number
of the massless fields in three dimension within the Kaluza-Klein point
of view. In the next section, we will see how the value of $\phi$ is
determined by one-loop quantum effect.

%%%%%%%%%%%%%%%%%%%%%%%%%%%%%%%%%%%%%%%%%%%%%%%%%%%%%%%%%%%%
\section{One-Loop Quantum Effect and Finite Temperature
}
%%%%%%%%%%%%%%%%%%%%%%%%%%%%%%%%%%%%%%%%%%%%%%%%%%%%%%%%%%%%
In order to calculate the one-loop effective potential for the model, we
use a covariant derivative with constant gauge fields. For instance,
one finds the expression of the effective potential for fundamental
fermion fields:
\begin{equation}
V_{eff}=-\frac{N_F}{\rm (Volume)} \ln\det({D\!\!\!\!/})\,,
\end{equation}
where
\begin{equation}
D_\mu=\partial_\mu-igA_y \delta_{y\mu}\,. 
\end{equation}
$N_F$ is the number of the fermion fields which belong to a fundamental
representation of $SU(2)$.

Of course, we also take care of the compactness of $S^1$ in the
calculation. Thus, the effective potential for the system under
consideration is given by \cite{8} :
\begin{equation}
V_{eff}=-\frac{2}{L^4\pi^2}\sum_{\ell=1}^\infty\frac{1}{\ell^4}
[2\cos (\ell\phi) + 1]
+ N_F\frac{4}{L^4\pi^2}\sum_{\ell=1}^\infty\frac{1}{\ell^4}
2\cos (\delta\ell)\cos\left(\ell\frac{\phi}{2}\right)\,,
\end{equation}
where $0\le\phi<4\pi$.
The first term comes from the contribution of $SU(2)$ gauge fields and
the other from fermions.

As Hosotani \cite{8} has pointed out, an absolute minimum of $V_{eff}$
depends on $\delta$; that is 
\begin{eqnarray}
\langle\phi\rangle&=&2\pi\quad {\rm for~} |\delta|\le \pi/2\\
 {\rm and} & &\nonumber\\
\langle\phi\rangle&=&0\quad {\rm for~} \pi/2\le|\delta|\le \pi\,.
\end{eqnarray}
When $|\delta|<\pi/2$, though the gauge symmetry is not
broken, the fermion masses in three dimensions become
\begin{equation}
\frac{1}{L}\left(2\pi n+\pi-\delta\right)\quad (n{\rm
:~integer})\,.
\end{equation}
Even if one imposes $\delta=0$, the result yields the antiperiodic
boundary condition i.e. there is no massless fermion in the three
dimensions. 

We can interpret this result as the breakdown of a $Z_2$
symmetry \cite{17}. In the model of this type for $SU(N)$, the breakdown
of a $Z_N$ symmetry may occur by introducing fermions.

Hosotani \cite{8} also pointed out that symmetry breakdown from $SU(2)$
to $U(1)$ may take a place with putting scalar bosons of fundamental
representation into the system. From now on, when we say ``symmetry''
it means those of $Z_N$ as well as $SU(N)$. Detailed issues, such as the
connection between gauge transformations and the boundary condition on
gauge fields, cannot be treated here. One finds an extensive discussions
in \cite{8,9,10}.

Now, we shall consider the finite temperature effect. In the one-loop
approximation, we perform a calculation of ``free energy'' at
temperature $T=\beta^{-1}$ similarly to the derivation of the
zero-temperature effective potential; only one difference is that we
must impose boundary conditions on fields with respect to the Euclidean
time coordinate $\tau$ ($0 \le \tau <\beta$), i.e., periodic for bosons
and antiperiodic for fermions \cite{14}. It is well known that these
conditions lead naive particle statistics.

At carrying out the calculation, the regularization is needed as we used
in the zero-temperature case; here, we simply drop the divergence
independent of $\phi$. Consequently, the free energy $\Omega$ is
obtained as follows:
\begin{eqnarray}
&
&\Omega=-\frac{2VL}{L^4\pi^2}\left[\frac{\pi^4}{30}\left(
\frac{L}{\beta}\right)^4+\frac{\pi^4}{90}+2\sum_{\ell=1}^\infty
\frac{\cos\ell\phi}{\ell^4}
\right.\nonumber \\
& &+\left.4\sum_{n=1}^\infty\sum_{\ell=1}^\infty
\frac{\cos\ell\phi}{\left(\frac{\beta^2}{L^2}n^2+\ell^2\right)^2}
+2\sum_{n=1}^\infty\sum_{\ell=1}^\infty
\frac{1}{\left(\frac{\beta^2}{L^2}n^2+\ell^2\right)^2}\right]
\nonumber \\
&
&+N_F\frac{4VL}{L^4\pi^2}\left[-\frac{7}{8}\frac{\pi^4}{90}\left(
\frac{L}{\beta}\right)^4+\sum_{\ell=1}^\infty
\frac{1}{\ell^4}\cos(\delta\ell)\cos\left(\ell\frac{\phi}{2}\right)
\right.\nonumber \\
& &+\left.2\sum_{n=1}^\infty\sum_{\ell=1}^\infty
\frac{(-1)^n}{\left(\frac{\beta^2}{L^2}n^2+\ell^2\right)^2}
\cos(\delta\ell)\cos\left(\ell\frac{\phi}{2}\right)\right]\,,
\label{FE}
\end{eqnarray}
where $V$ is the two dimensional volume of the system.

At a glance, one finds a peculiarity of the symmetry breaking
mechanism; the expression for the free energy of the system consists of
the trigonometrical functions. It is expected that possible extrema of
the free energy locate at the peculiar values of $\phi$ (i.e.,
$0, \pi, 2\pi$ etc.) since the coefficient of the
``high-frequency-wave'' functions are well suppressed by a factor $\sim
\ell^{-D}$ or $n^{-D}$, where $D$ is the total dimensions of the
space-time.

Now we consider the high temperature limit $(\beta/L\rightarrow 0)$.
Using asymptotic forms of the following numerical series,
\begin{eqnarray}
\sum_{n=-\infty}^\infty\frac{1}{(n^2+a^2)^2}&=&\frac{\pi}{2}\frac{1}{a^3}
+O(e^{-2\pi a}/a^3)\\
\sum_{n=-\infty}^\infty\frac{(-1)^n}{(n^2+a^2)^2}&=&O(e^{-\pi a}/a^3)
\qquad (a\gg 1)\,,
\end{eqnarray}
we can approximate the free energy (\ref{FE}) at $\beta/L\ll 1$:
\begin{equation}
\Omega\sim-VL\frac{\pi^2}{90}\left(2\times
3+\frac{7}{8}\times 4\times N_F\right)\frac{1}{\beta^4}
-\frac{2VL}{L^3\beta} \left[2\sum_{\ell=1}^\infty
\frac{\cos(\ell\phi)}{\ell^3}+\zeta(3)\right]\,.
\end{equation}

Accordingly, we observe that the $\phi$-dependent part of the free
energy (while the rest is pure radiation) which comes from gauge fields
becomes dominant in the high temperature limit; it has degenerate minima
at $\phi=0$ and $2\pi$.

Though the fermionic contribution becomes much suppressed at high
temperature, a careful evaluation reveals that the location of the
minima is not changed from the values at zero temperature; at high
temperature the contribution from ``high frequency'' trigonometrical
functions are exponentially suppressed. 

After all, it is found that
there is no change of symmetry pattern, i.e., symmetry restoration or
breakdown (of $Z_2$) in this system does not occur by raising
temperature. 

If one wants to make a system whose symmetry
depends on temperature, we must manage the combination of bosons and
fermions with various representations in general gauge theories. For
example, we suppose an $SU(2)$ gauge theory with $N_F$ fermions and
$N_S$ scalar bosons which are both in fundamental representation.
Further, if we impose $\delta\sim\pi/2$ as boundary conditions on both
fields and set $N_S=2N_F\gg 1$, one finds the gauge symmetry breaking by
raising temperature of the system.

To summarize, the extrema of the free energy, or effective potential in
terms of $\phi$ is almost insensitive to finite temperature effect. To
make the model whose symmetry pattern depends on temperature, we must
make use of the difference of statistics between bosons and fermions.
These properties are held in models with Hosotani's mechanism.

%%%%%%%%%%%%%%%%%%%%%%%%%%%%%%%%%%%%%%%%%%%%%%%%%%%%%%%%%%%%
\section{Effect of Finite Density and the Degenerate Fermions}
%%%%%%%%%%%%%%%%%%%%%%%%%%%%%%%%%%%%%%%%%%%%%%%%%%%%%%%%%%%%
In this section, we investigate a finite density effect on the
Hosotani's model which has been treated previously. For this purpose,
we employ the chemical potential for fermions; at the same time, the
free energy is extended to the ``thermodynamic potential''. A
prescription of deriving the expression for the thermodynamic potential
at one-loop level is already written down \cite{18}; in a course of the
calculation the covariant derivative need to be modified as
\begin{equation}
D_0=\partial_0\rightarrow \partial_0-iA_0\,,\quad
{\rm where~} A_0=-i\mu\,.
\end{equation}
Thus we only carry out the usual one-loop calculation as there is a
constant but imaginary $U(1)$ gauge potential in zeroth component. It
is interesting to find a similarity in the one-loop calculation for
compact space and the periodic time direction.

Finite density effect on various symmetry breakdown mechanisms have
been studied extensively \cite{15}. In the Kaluza-Klein context, the
thermodynamic potential with $\mu\ne 0$ was discussed by the present
author \cite{19}.

The calculation technique is parallel to the one used in \cite{19}. The
expression is given in Appendix. Here, we consider the strongly
degenerate fermi gas, i.e., in the situation of $\mu\ne 0$ and
$T\rightarrow 0$.
For simplicity, we shall give the expression for the thermodynamic
potential of fermions in fundamental representation of $SU(2)$ with
periodic boundary condition ($\delta=0$) on $S^1$.

Within the Kaluza-Klein point of view, it is interesting to study the
case that $\mu$ is less than, or at most nearly equals to $2\pi/L$.
Accordingly, we only consider the case that $0<\mu L<\pi$ here.

The thermodynamic potential, or the effective potential of $\phi$ with
$\mu\ne 0$, is given by: 
\begin{eqnarray}
\Omega&=&N_FVL\frac{1}{L^4\pi^2}\left[\frac{1}{6}\left\{
2\pi^2\left(\frac{\phi}{2}-\pi\right)^2-
\left(\frac{\phi}{2}-\pi\right)^4-\frac{7}{15}\pi^4\right\}\right.
\nonumber
\\ & &-\left.\frac{\pi}{3}\left(\frac{\phi}{2}-\mu L\right)^2
(\phi+\mu L)\right]\nonumber \\
& & {\rm for~} 0<\phi/2\le\mu L\,,
\label{fa}
\end{eqnarray}
\begin{eqnarray}
\Omega&=&N_FVL\frac{1}{L^4\pi^2}\left[\frac{1}{6}\left\{
2\pi^2\left(\frac{\phi}{2}-\pi\right)^2-
\left(\frac{\phi}{2}-\pi\right)^4-\frac{7}{15}\pi^4\right\}\right.
\nonumber \\
& & {\rm for~} \mu L<\phi/2\le 2\pi-\mu L\,,
\label{fb}
\end{eqnarray}
\begin{eqnarray}
\Omega&=&N_FVL\frac{1}{L^4\pi^2}\left[\frac{1}{6}\left\{
2\pi^2\left(\frac{\phi}{2}-\pi\right)^2-
\left(\frac{\phi}{2}-\pi\right)^4-\frac{7}{15}\pi^4\right\}\right.
\nonumber
\\ & &-\left.\frac{\pi}{3}\left(2\pi-\frac{\phi}{2}-\mu L\right)^2
(4\pi-\phi+\mu L)\right]\nonumber \\
& & {\rm for~} 2\pi-\mu L<\phi/2\le 2\pi\,,
\label{fc}
\end{eqnarray}

An absolute minimum of $\Omega$ is given by $\phi/2 = \pi$ for
$\mu L<2^{-1/3}\pi$ and $\phi/2=0$ for $2^{-1/3}\pi<L<\pi$. For 
$\mu L =2^{-1/3}\pi$ these two minima are degenerate. 

One can find that the non-zero density of fermion may restore the
symmetry. It is remarkable that finite density effect is much important
for symmetry breaking while finite temperature has almost no effect on
it.

In the region $\mu L>\pi$, for sufficient large $\mu L$ ($>1.9 \pi$)
the point $\phi/2 =\pi$ becomes an absolute minimum again; moreover,
for $\mu L > 2\pi$, new minima whose locations depend on $(\mu L)$ even
appear. Furthermore, in the region $\mu L \gg 2\pi$, finding of the
vacuum requires rather complicated investigations.

The analysis of symmetry pattern will be more complicate if we consider
the model with realistic larger gauge group.

%%%%%%%%%%%%%%%%%%%%%%%%%%%%%%%%%%%%%%%%%%%%%%%%%%%%%%%%%%%%
\section{Discussion}
%%%%%%%%%%%%%%%%%%%%%%%%%%%%%%%%%%%%%%%%%%%%%%%%%%%%%%%%%%%%
In this paper, we calculate the one-loop effective potential at finite
temperature and density, or the thermodynamic potential, for the
simple model suggested by Hosotani \cite{8} and investigate its symmetry
breaking pattern. It turns out that the degenerate fermions have
much influence on the symmetry breaking through the thermodynamic
potential, while the finite temperature only gives the correction which
has almost the same shape as the effective potential at $T = 0$.

Therefore, we should look for the evidence of the compact dimensions
inside the cold fermion star as well as near the black holes \cite{20}!
Actually, however, only ``theoretical'' searches can be done for simple
models, one of which is examined in this paper. In order to use the
Hosotani's mechanism in unified theories, we must correctly of course
take into account the gravitational effects and, realistic gauge group
and compact manifold.

$S^1$, the multiply-connected space, is a rather exceptional one. In
the context of compactified string models, the manifold $M$ written as
$M= K/H$ where $K$ is a simply-connected manifold and H is a finite
group is often considered; then $\Pi_1(K/H)= H$. In this case, a Wilson
loop can be regarded as an element of a finite group $H$, while
$\Pi_1(S^1)=Z$ is an infinite group. One-loop vacuum energy has been
calculated by Evans and Ovrut \cite{11} for some simple cases such
as $M =S^3/Z_2$ . The method of the calculation is fairly different from
the one for the model in the present paper, because symmetry of the
finite group restricts the eigenmode on $M$. It is also important to
consider the finite temperature and density effect on such models.
Because the temperature and/or density may break the supersymmetry, it
is of interest to study a model with various particle content.

On the other hand, the presence of gauge fields implies the existence of
conserved quantities in the system. Moreover, the asymptotic-free
nature of non-Abelian gauge theory imposes the ``colorlessness'' of
the system. The effect have been considered in the investigation of the
``quark-gluon plasma'' \cite{21}. The colorless partition function can
be obtained by averaging the naive partition function in terms of the
zeroth component of the vacuum gauge fields; i.e., the covariant
derivative $D_0 =\partial_0-i{\bf A}_0$ is employed to derive the
one-loop result. It is naturally expected that the global color
confinement and the non-trivial Wilson loop element on
multiply-conected space are influenced by each other as is in the case
of $\mu\ne 0$.

Imposing the color singlet condition may be of great relevance not only
to the dynamics in the very early universe as is pointed out in
\cite{22} but also the Kaluza-Klein cosmology, on account of shrinking
extra dimensions.

However, since matter fields play both roles to break the symmetry and
to spoil the confinement, the matter content of the theory must be
specified within certain extent for a detailed analysis. Therefore
we have left the discussion on the effect of the ``colorlessness''
condition with the symmetry breaking by Wilson loops until more
realistic models appear.

%%%%%%%%%%%%%%%%%%%%%%%%%%%%%%%%%%
\section*{Acknowledgement}
%%%%%%%%%%%%%%%%%%%%%%%%%%%%%%%%%%
The author would like to thank M. Hosoda for reading this manuscript and
useful comments.

%%%%%%%%%%%%%%%%%%%%%%%%%%%%%%%%%%
\section*{Appendix}
%%%%%%%%%%%%%%%%%%%%%%%%%%%%%%%%%%
Here, we shall briefly show the calculation of the thermodynamic
potential treated in the text. In the space $R^d\times S^1$, the
thermodynamic potential for $SU(2)$ doublet fermions with $\mu\ne 0$ and
$\delta=0$ is formally expressed as follows:
\begin{eqnarray}
& &\frac{4N_F}{\beta} \frac{V}{(2\pi)^d} \int_0^\infty dt\, t^{-1}
\int d^d k \sum_{n=-\infty}^\infty \sum_{\ell=-\infty}^\infty\nonumber
\\& &\cdot {\rm
Re}\left[\exp\left\{-t\left(k^2+\left(\frac{(2n+1)\pi}{\beta}
+i\mu\right)^2+\left(\frac{2\pi\ell+\phi/2}{L}\right)^2\right)
\right\}\right]\nonumber \\
& &(0 \le \phi/2 < 2\pi)\,. 
\end{eqnarray}
After dropping the divergence which is independent of $\phi$, it
becomes:
\begin{eqnarray}
& &\Omega=4N_F VL\left[\frac{1}{\pi^{\frac{d+2}{2}}} \Gamma\left(
\frac{d+2}{2}\right)\cdot 2\sum_{\ell=1}^\infty\frac{1}{(L\ell)^{d+2}}
\cos\left(\ell\frac{\phi}{2}\right)\right.\nonumber \\
& &\quad+\frac{1}{L(4\pi)^{d/2}}
4\sum_{n=1}^\infty (-1)^n\cosh(\mu\beta n)\left\{
\left(\frac{\phi}{L\beta n}\right)^{\frac{d+1}{2}}K_{\frac{d+1}{2}}
\left(\frac{\beta\phi}{2L}n\right)
\right.\nonumber \\
& &\quad+\sum_{\ell=1}^\infty\left[
\left(\frac{2(2\pi\ell+\phi/2)}{L\beta
n}\right)^{\frac{d+1}{2}}K_{\frac{d+1}{2}}
\left(\beta n\frac{2\pi\ell+\phi/2}{L}\right)
\right.\nonumber\\
& &\quad\quad\qquad+\left.\left.\left.
\left(\frac{2(2\pi\ell-\phi/2)}{L\beta
n}\right)^{\frac{d+1}{2}}K_{\frac{d+1}{2}}
\left(\beta n\frac{2\pi\ell-\phi/2}{L}\right)
\right]\right\}\right].
\end{eqnarray}
(Here, we used some identities which can be found in \cite{19}.)

In order to find the expression in the case of
$T\rightarrow 0$ ($\beta\rightarrow\infty$), we use an integral
representation of the modified Bessel function:
\begin{equation}
K_\nu(z)=\frac{\sqrt{\pi}(z/2)^\nu}{\Gamma(\nu+1/2)}\int_1^\infty
e^{-zx} (x^2-1)^{\nu-1/2} dx\,.
\end{equation}
Using this, we can perform the summation over $n$ to obtain:
\begin{eqnarray}
& &\Omega=4N_F VL\left[\frac{1}{\pi^{\frac{d+2}{2}}} \Gamma\left(
\frac{d+2}{2}\right)\cdot 2\sum_{\ell=1}^\infty\frac{1}{(L\ell)^{d+2}}
\cos\left(\ell\frac{\phi}{2}\right)\right.\nonumber \\
& &\quad-\frac{1}{L}\frac{1}{(4\pi)^{d/2}}
\frac{1}{\Gamma\left(\frac{d+2}{2}\right)}\left\{
\left(\frac{\phi}{2L}\right)^{d+1}\int_1^\infty dx (x^2-1)^{d/2}
\right.\nonumber \\
&
&\quad\cdot\left(\frac{1}{\exp\left[\beta
\left(\frac{\phi}{2L}x-\mu\right)+1\right]}+(\mu\rightarrow-\mu)\right)
\nonumber \\
& &\quad+\sum_{\ell=0}^\infty\left\{\left[
\left(\frac{2\pi\ell+\phi/2}{L}\right)^{d+1}
\int_1^\infty dx (x^2-1)^{d/2}\right.
\right.\nonumber\\
& &\cdot\left.\left.\left.\left.
\left\{\frac{1}{\exp\left[\beta
\left(\frac{2\pi\ell+\phi/2}{L}x-\mu\right)+1\right]}+(\mu\rightarrow-\mu)
\right\}\right]+[\phi\rightarrow-\phi]\right\}\right\}\right].
\end{eqnarray}
In the limit of $\beta\rightarrow\infty$,
\begin{equation}
\frac{1}{e^{\beta
x}+1}\stackrel{\beta\rightarrow\infty}{\longrightarrow} \theta(-x)
\end{equation}
is recognized, and at the same time an infinite series becomes a finite
sum over $\ell$.

Soon we can find the expressions (\ref{fa}, \ref{fb}, \ref{fc}) in the
text when $d = 2$, $0<\mu L<\pi$.

%%%%%%%%%%%%%%%%%%%%%%%%%%%%%%%%%%%%%%%%%%%%%% 

%%%%%%%%%%%%%%%%%%%%%%%%%%%%%%%%%%%%%%%%%%%%%

\begin{thebibliography}{99}
%%%%%%%%%%%%%%%%%%%%%%%%%%%%%%%%%%%%%%%%%%%%%% 
\bibitem{1} Selected papers can be found in: J. Schwarz (ed.),
{\it Superstrings} (Singapore: World Scientific 1985).
\bibitem{2} M. Green and J. Schwarz, Phys. Lett. {\bf B149} (1984) 119.
\bibitem{3} D. Gross, J. Harvey, E. Martinec and R. Rohm, Phys. Rev.
Lett. {\bf 54} (1985) 502; Nucl. Phys. {\bf B256} (1985) 253; Nucl.
Phys. {\bf B267} (1986) 75.
\bibitem{4} J. Ellis, K. Enqvist, D. V. Nanopoulos and F. Zwirner, Nucl.
Phys. {\bf B276} (1986) 14.
\bibitem{5} Th. Kaluza, Sitzungsber. Preuss. Akad. Wiss., Berlin, Math.
Phys. {\bf K1} (1921) 966; O. Klein: Z. Phys. {\bf 37} (1926) 895; E.
Witten, Nucl. Phys. {\bf B186} (1981) 412, and references therein.
\bibitem{6} P. Candelas, G. Horowitz, A. Strominger and E. Witten, Nucl.
Phys. {\bf B258} (1985) 46.
\bibitem{7} E. Witten, Nucl. Phys. {\bf B258} (1985) 75; J. Breit, B.
Ovrut and G. Segr\`e, Phys. Lett. {\bf B158} (1985) 33; A. Sen, Phys.
Rev. Lett. {\bf 55} (1985) 33.
\bibitem{8} Y. Hosotani, Phys. Lett. {\bf B126} (1983) 309.
\bibitem{9} D. Toms, Phys. Lett. {\bf B126} (1983) 445.
\bibitem{10} S. Midorikawa and M. Tomiya, Preprint INS-Rep.-597 (July
1986).
\bibitem{11} M. Evans and B. A. Ovrut, Phys. Lett. {\bf B174} (1986) 63;
B.A. Ovrut, Prog. Theor. Phys. Suppl. {\bf 86} (1986) 185.
\bibitem{12} V.B. Svetovo\v{i} and N.G. Khariton, Sov. J. Nucl. Phys.
{\bf 43} (2) (1986) 280.
\bibitem{13} For example, see Nucl. Phys. {\bf B252}, Nos. 1, 2 (1985).
\bibitem{14} D. A. Kirzhnits and A. D. Linde, Phys. Lett. {\bf B42}
(1972) 471; S. Weinberg, Phys. Rev. {\bf D9} (1974) 3357; L. Dolan and
R. Jackiw, Phys. Rev. {\bf D9} (1974) 3320; A. D. Linde, Rep. Prog.
Phys. {\bf 42} (1979) 389.
\bibitem{15} D. Bailin and A. Love, Nucl. Phys. {\bf B226} (1983) 493;
{\bf B233} (1984) 204; A. Love and S. J. Stow, Nucl. Phys. {\bf B243}
(1984) 537; {\bf B257} [FS14] (1985) 271.
\bibitem{16} F. S. Accetta and E.W. Kolb, Phys. Rev. {\bf D34} (1986)
1798, and references therein.
\bibitem{17} N. Weiss, Phys. Rev. {\bf D25} (1982) 2667.
\bibitem{18} A. Actor, Nucl. Phys. {\bf B265} [FS15] (1986) 689, and
references therein.
\bibitem{19} K. Shiraishi, Prog. Theor. Phys. {\bf 77} (1987) 1253.
\bibitem{20} A. Davidson and D.A. Owen, Phys. Lett. {\bf B155} (1985)
247; M. Yoshimura, Preprint KEK-TH 116 (1985); Phys. Rev. {\bf D34}
(1986) 1021; S. Midorikawa and M. Tomiya, Preprint INS-Rep. 603 (1986).
\bibitem{21} B. M\"uller, {\it The physics of the quark-gluon plasma}
(Berlin, Heidelberg, New York, Springer 1985); H. T. Elze and W.
Gleiner, Phys. Rev. {\bf A33} (1986) 1879, and references therein.
\bibitem{22} B. S. K. Skagerstam, Phys. Lett. {\bf B133} (1983) 419.
\end{thebibliography}
\end{document}